# Research on Anonymization and De-anonymization in the Bitcoin System

QingChun ShenTu[12*], JianPing Yu [1]

[1] ATR Defense Science & Technology Lab., Shenzhen University, Shenzhen, China
[2] Bitbank Research Labs
[*] unshadowster@gmail.com

**Abstract:** The Bitcoin system is an anonymous, decentralized crypto-currency. There are some deanonymizating techniques to cluster Bitcoin addresses and to map them to users' identifications in the two research directions of Analysis of Transaction Chain (ATC) and Analysis of Bitcoin Protocol and Network (ABPN). Nowadays, there are also some anonymization methods such as coin-mixing and transaction remote release (TRR) to cover the relationship between Bitcoin address and the user. This paper studies anonymization and de-anonymization technologies and proposes some directions for further research.

1. **Introduction**

Bitcoin is a decentralized crypto-currency that was introduced by Nakamoto [1] in 2008, and was deployed in January 2009. Bitcoin has several characteristics including peer-to-peer protocols, decentralized production of Bitcoins by the proof of work (PoW) protocol, the prevention of double spending by transparent transactions, pseudo-anonymity, and personal privacy protection, which have made Bitcoin increasingly popular. Its market value reached at peak of $12 billion in 2013.

In 2008, Pfitzmann [2] gave the anonymity a definition, and anonymity of a subject means that the subject is not identifiable within a set of subjects, denoted as the anonymity set. In the Bitcoin network, Bitcoin addresses act as user accounts. Generally speaking, the aim of anonymization is to prevent attackers from discovering the relationship between Bitcoin addresses and real or virtual user identity information through the Bitcoin network and the blockchain used to record transactions. Conversely, deanonymization is the uncovering of the relationship between the Bitcoin address and the user.

Anonymity in the Bitcoin system is based on the following facts: (1) Bitcoin address cannot be mapped to the real identity; (2) Bitcoin transactions don't contain any personal information; (3) The new transactions are spread radially, thus the sender's IP address will not be exposed. However, the weakness of Bitcoin anonymity is reflected in the following areas: (1) The real-name authentication mechanism helps Bitcoin service providers to find the addresses that ever deposited and withdrew; (2) Bitcoin address exposed on the internet can be related to its owner; (3) The chain of transactions is transparent and traceable; (4) Gathering some or all inputs when sending Bitcoins to others, which may expose other addresses of the sender; (5) The change address of transactions could be classified by attackers to the sender.

The current study on Bitcoin deanonymization focuses on two methods. One is the Analysis of the Transaction Chain (ATC), which is to obtain transactions from public blockchain data, to classify Bitcoin addresses based on the weakness of Bitcoin anonymity [3-8], and to relate Bitcoin addresses to personal identities [7-8]. The other method is the Analysis of the Bitcoin Protocol and Network (ABPN). This makes use of the spreading characteristics of Bitcoin transactions to deduce the source IP address of a new transaction [9-13], and known attacks include the Bitcoin protocol sniffer, Sybil attack [10-12], fake Bitcoin nodes [13]. The method to combat ATC attacks is coin-mixing [14-30], and methods against APBN attacks are The Onion Router (TOR) [32], The Invisible Internet (I2P) [33] and Transaction Remote Release (TRR) [34].

## 2. ATC and coin-mixing

### 2.1. Transactions and the transaction chain

A Bitcoin address is an account on the Bitcoin network, which corresponds to a bank account in conventional currency systems. A Bitcoin address is generated by double hashing a public key. Only the user who owns the corresponding private key can make use of Bitcoins lodged at this address.

The Bitcoin system has a public ledger that stores transfer records rather than the balance of every Bitcoin address. A transfer record is known as a Bitcoin transaction [35], and it includes the transfer time, inputs, outputs, amounts, and signatures. Bitcoin transactions with their related inputs and outputs enter into the transaction chain, as illustrated in Fig. 1.

Every transaction has a unique identification (ID). Each input is connected with the output of the previous transaction so that the input address of a transaction can be obtained through the output address of the previous transaction. All inputs should not be used within all existing transactions because this will prevent the successful verification of the transaction. Signatures aim to prove that an input amount belongs to the sender because only the private key owner can sign the transaction properly.

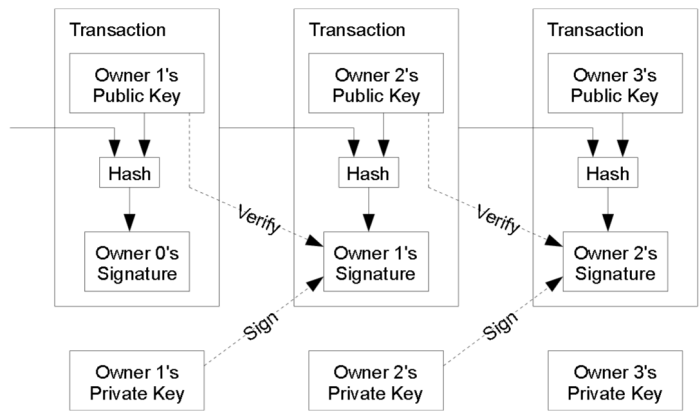

***Fig. 1** Bitcoin transactions and the chain of transactions*

Transactions data are stored in the Bitcoin blockchain [36] composed of a lot of blocks, each block is like a page in the public ledger, recording a block header and all the transactions released in the Bitcoin system during the latest ten minutes. There is an item denotes as the previous hash value in the block header, it means, a block includes the previous hash value, and the hash value of this block is also included by the next block, thus, all the blocks enter into a chain of blocks, denotes as blockchain. Nakamoto adopted PoW[1] to ensure generating blocks is decentralized, regular and secure. The miner generating the blocks could be rewarded some Bitcoins, which is 50 BTC in the first four years, and halves every four years, and it is 25 BTC nowadays. In addition, all transaction fees of transactions within a block, which equal to the balance of the output amount of these transactions minus the input amount, are a part of the block reward.

*2.2. ATC*

ATC is able to cluster many Bitcoin addresses to one user, or find clue to connect Bitcoin addresses with real or virtual user identity information. Such methods are transaction and user networks [3], taint analysis [4], amount analysis and timing sequence.

Each vertex represents a user and each directed edge between a source and a target represents a transaction with a value in Bitcoins and a timestamp, which denotes as the user network. As shown in Fig.2 (a). Similarly, each vertex represents a transaction and each directed edge between a source and a target represents an output of the transaction with a value in Bitcoins and a timestamp, which denotes as the transaction network. As shown in Fig.2 (a). When starting from a vertex, after a series of transactions, and then back to the vertex, called the transaction chain closure.

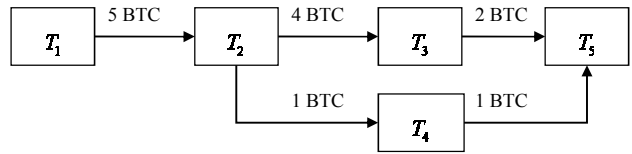

(a) The transaction network

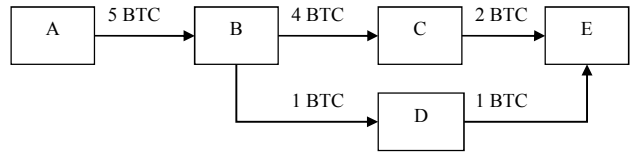

(b) the user network

*Fig. 2* *The transaction and user networks*

*Taint analysis* is a service provided by Blockchain.info, which is used to calculate the percentage of Bitcoins in an address from another address. Obviously, taint analysis is premised on that these two addresses are related to each other in a transaction chain. As shown in Fig.2 (b), the degree of taint of C and E, D and E are 2/3, 1/3 respectively.

*Amount analysis* is another method to deduce the relationship between the inputs and the outputs within a CoinJoin transaction, according to the characteristic that the input amount of a Bitcoin transaction equals to the output amount plus transaction fee. There is a CoinJoin transaction shown in Fig.4, known condition that coin-mixing fee is 1%, we could deduce that the $0^{th}$ input corresponds to the $0^{th}$ output, and the $1^{st}$ input corresponds to the $1^{st}$ output, and the $2^{nd}$ and $3^{rd}$ inputs correspond to the $2^{nd}$ output, and then the $3^{rd}$ output is coin-mixing fee.

*Timing sequence*: after the coin-mixing service provider (mixer) completes the coin-mixing operation he should return Bitcoins to users within a specific time. Hence, the attacker could estimate that the transaction returning Bitcoins were within a block range, and interacting with amount analysis so as to reduce users' possible output addresses to a lesser extent.

Reid [3] analyzed the transaction chain in Bitcoin system and cluster addresses using the transaction and user networks mentioned above in 2011. They studied the event that 25,000 BTC belonging to the Slush Pool was stolen in 2011, and demonstrated how to trace transaction through transaction chain and external information, and they gained some useful clues. Ober [5] and Ron [6] analyzed the Bitcoin blockchain, and studied the amount change of Bitcoins in all Bitcoin addresses, and studied the procedure how users acquired and spent their Bitcoins, and how they transferred Bitcoins within multiple addresses in order to protect privacy. Ron [6] also traced 364 transactions which amount is larger than 50,000 BTC, and found that they were related to a transaction with 90,000 BTC in 2010.

Androulaki [7] simulated Bitcoin payment scenarios in a university in 2012 and estimated the anonymity degree of Bitcoin quantitatively through two heuristic rules, gathering some or all inputs in a transaction and change address. They found that even the receiver generates a new address each time when receiving Bitcoins, 40% of users could be found the real identify. Meiklejohn [8] clustered addresses within the Bitcoin blockchain based on the same heuristic rules and identified 1.9 million addresses, which accounted for 16% of all addresses at that time. He also connected these addresses with addresses of same Bitcoin service provider and found 500,000 addresses that Mt.Gox used before and 250,000 addresses that Silk Road used before.

The analysis methods mentioned above are not always able to find users' identities. It is partly because of being short of the corresponding relationship between real or virtual identity and Bitcoin addresses; another reason is that the heuristic rules would bring about some mistakes such as coin-mixing inputs making the rule of gathering inputs get wrong inputs, identifying wrong change address, other users' addresses including in a transaction chain closure. In spite of shortages, the above analysis methods can provide a variety of valuable clues, if it is relative to users' identities of Bitcoin service providers, and then there is a high probability to find the true identity of a specific Bitcoin address.

*2.3. Coin-Mixing*

For the ATC attacks, people obfuscate the transaction chain, and separate the corresponding relationship between the input and output of a Bitcoin transaction and even hide the amount of transaction, which denoted as coin-mixing.

Current study regarding coin-mixing focuses on three directions. The first is the use of centralized coin-mixing, such as DarkWallet [14], BitcoinFog.com, BitLaundry.com, and Blockchain.info, where CoinJoin [15], Mixcoin [16], and coin-mixing algorithms based on blind signature such as BlindCoin [17], RSA Coin-Mixing [18] and Blind-Mixing [19] are representative studies. The second regards decentralized protocols such as CoinSwap [20], Fair Exchange [21], XIM [22], CoinShuffle [23] and CoinParty [24], which are compatible with the Bitcoin protocol, where Bitcoins are mixed through decentralized protocols, and no trust is required among users with no possibility of currency loss. The last direction regards novel coin-mixing technologies such as blind signature transaction [25], Zerocoin [26], Zerocash [27], Pinocchio [28], CryptoNote [29] and SideChain [30], which are not compatible with the Bitcoin Protocol, and must be applied in new blockchains.

*2.3.1 Centralized coin-mixing*: research regarding Bitcoin coin-mixing originated from the CoinJoin anonymization method proposed by Gmaxwell [15]. General Bitcoin transactions have 1 to 2 inputs and 1 to 2 outputs, so it is easy to analyze the transferring path of Bitcoins. The CoinJoin transction combines many inputs and outputs and put them into a single transaction so that an input in the CoinJoin transaction is difficult to be corresponded to an output. Suppose that the numbers of inputs and outputs within a transaction are all N, and each input corresponds to a output respectively, and given event A is that corresponds a certain input to a certain output. When N = 10, then the possibility of A $P(A) = 0.1$; after 5 coin-mixing activities, $P(A)=10^{-5}$, it seems CoinJoin transaction is the most simple and most effective anonymization method.

According to above principle, centralized coin-mixing providers receive Bitcoins from users through website and perform coin-mixing activities. They have some countermeasures against amount analysis: ① each user inputs same amount; ② random coin-mixing fee, attackers cannot estimate output amount accurately; ③ extend coin-mixing path and increase the frequency of coin-mixing, so that attackers are not able to perform amount analysis within a single CoinJoin transaction; ④ reduce the amount of coin-mixing, such as 0.1 to 5 BTC each time; ⑤ receive Bitcoins from multiple addresses.

Methods against the timing sequence: ① buy more time to mix coins, with higher probability to be obfuscated with other normal transactions ; ② make the time when the mixers send Bitcoins back to users random within a long range, such as from 2 days to 7 days, attackers cannot estimate the time accurately.

Method against the user and transaction networks is make use of shared wallet, that is, the mixers receive Bitcoins through address A, however, send Bitcoins to users through address B. Thus, the input and output addresses of a certain user are no relation in the transaction chain.

The mixers hide users' IP addresses through TOR or I2P, and receive Bitcoins from users, and then perform coin-mixing, therefore develop the anonymity of the Bitcoin system. However, there are two serious weaknesses of centralized coin-mixing: ① mixers must know a user's input address and output addresses, and hence cannot provide true anonymity for users; ② users must trust mixers and send Bitcoins to mixers in advance, so users face with the risk of currency loss.

A blind digital signature represents a condition where a signer signs the digest of a message while the content of the message is unknown to the signer. In 1982, Chaum [31] first proposed to implement an anonymous e-Cash system based on the use of blind signatures, which was intended to protect the anonymity of a sender unconditionally. In 2015, BlindCoin [17] was proposed based on bilinear groups to make centralized coin-mixing more anonymous; however, it may be deanonymized because that it uses a public log to reach accountability, which may reveal the deposit time and the withdrawal time. In July 2015, Wu [18] also proposed a blind coin-mixing algorithm based on RSA (RSA Coin-Mixing), however, a user's Bitcoins may be falsely claimed by another. In 2015, Shentu et al [19] adopted elliptic curve cryptography (ECC) and blind signature, and propose a Blind-Mixing scheme, thus improve the performance of ECC blind signature. The scheme also uses the private key from the user' input address to sign the deposit voucher, thus the Bitcoins sent to the mixer cannot be falsely claimed by another.

As for the users' risk of currency loss, the present studies such as Mixcoin and BlindCoin intend to make centralized coin-mixing reach accountability. However, they could not prevent Bitcoins loss when the mixers were gone. Therefore, it is also an important subject.

*2.3.2 Decentralized coin-mixing protocols*: because of the risk of currency loss and the application of the secure multi-party computation protocol (SMC) [37] in the Bitcoin [38, 39, 40], there are some decentralized coin-mixing protocols, and no trust is required among users with no possibility of currency loss.

Fair Exchange [21] was proposed by Barber, which is a two-party Bitcoin exchange protocol, and the two parties exchange their Bitcoins without any trust between each other using Bitcoin script and three Bitcoin transactions including commit transaction, refund transaction and claim transaction.

XIM [22] is a two-party Bitcoin exchange protocol resisting Sybil attacks and DOS attacks, no third party is required. Bissias invents a method finding coin-mixing partners in the Bitcoin blockchain, and adopts Fair Exchange as exchange protocol, and prevent Sybil attacks and DOS attacks through transaction fee.

CoinShuffle [23] is a decentralized multi-party coin-mixing protocol, and communicates with other partners using anonymous group communication protocol Dissent [41] to ensure anonymity. When coin-mixing, all output addresses from all users would be changed the order randomly, and no one could know the corresponding relationship between users and their output addresses. At last, a CoinJoin transaction with multi-input and multi-output will be generated and released.

CoinParty [24] is also as multi-party coin-mixing protocol based on SMC, and there is a centralized server acting as a communication platform to gathering coin-mixing requirements and to transfer messages among users. When coin-mixing, an escrow address is generated and controlled by all coin-mixing partners, and all users send Bitcoins to this address, and then all output addresses are obfuscated through the similar method as CoinShuffle does, and at last a CoinJoin transaction, which input address is the escrow address, is generated and released.

*2.3.3 Decentralized coin-mixing Protocols in blockchain*: are not compatible with the Bitcoin Protocol, and must be applied in new blockchains. Ladd [25] introduced blind signature to Bitcoin transaction the first time, and proposed the idea of blind signature transaction, which means that the sender intents to send Bitcoins to a group of receivers, and then he signs and releases the transaction, however he cannot connect receivers with their addresses. Blind signature transaction is a good theoretical exploration, fusing the blind signature, zero knowledge proof and Paillier public-key system.

Amount analysis is an import ATC method, however, the SideChain [30] developed by BlockStream has a kind of secret transaction using Pederson Commitment to replace amount, where the commitment stands for an Bitcoin amount within a certain range such as from 0 to 2 BTC but anyone don't know exact number except the private owner. Even so, the commitment could be verified by the third party and ensure the owner not to spend more money. Thus, Pederson Commitment renders the amount analysis ineffective.

Ring signature was first introduced in CryptoNote [29], here are features: ① the signer selects any user's public keys to take part in signing, no notify required; ② unforgeablity, attackers don't know any member's private key, so they cannot forge signature; ③ unconditional anonymity, in case attackers gain all possible private keys, however, the possibility of the signer being recognized is not more than $1/n$, where n indicates the number of possible signers. Some crypto-currencies bases on CryptoNote such as Bytecoin [42] and DarkNetSpace [43] used ring signature to hide the sender, equally performed a coin-mixing activity.

Stealth address was also first introduced by CryptoNote, which is originated from the Diffie-Hellman encryption key exchange protocol based on Elliptic Curve (ECDH). The receiver makes a special address public, where the address is called Stealth address, the sender transfers Bitcoins to the address with a one-time public key, and attackers are unable to find any transactions according to the address. However, the receiver could calculate the correct receiving address and the corresponding private key, and gains the Bitcoins transferred by the sender. DarkWallet [14] and BitShares [44] are also using this technology.

Zerocoin [26] is a decentralized coin-mixing protocol built on new blockchain, including Mint and Spent these two kinds of operations. Mint operation transforms Bitcoins to Zerocoins, and Spent operation transfers Zerocoins in the Zerocoin blockchain or exchanges Zerocoins to Bitcoins. Zerocoin hides the addresses of the sender and the receiver, however, some weaknesses exist: ① the man who defines the initiation data can acquire all Zerocoins; ② the performance is really not good, the size of zero knowledge in each transaction is larger than 45KB, and it needs 450 ms time to verify; ③ Can't split the amount, and cannot be used to make payment.

Zerocash [27] is the improved version, here are features: ① the initiation data being defined by users; ② the size of zero knowledge being decreased 97.7%; ③ the verification time being reduced by 98.6%; ④ it could be used to pay; ⑤ input amount could be changed or hidden. Another improved version Pinocchio [28] reduced the size of zero knowledge to 288 bytes, and the verification time is less than 10ms.

*2.3.4 Quantifying the anonymity*: in order to estimate the anonymity degree of coin-mixing transaction and coin-mixing system, we need to quantify the anonymity degree. Suppose a coin-mixing system with N members. When one of members sends a message, ideally, the probability of identifying him is 1/N. However, if attackers learn some knowledge through participating in and observing this system, the probability of some members being identified will increase. Diaz [45] used the following quantifying mode to calculate the anonymity degree.

$$d = \frac{H(X)}{H_M} = \frac{-\sum_{i=1}^{N} p_i * log_2(p_i)}{log_2(N)} \quad (1)$$

where H(X) indicates the entropy of attacked system, pi indicates the probability of the ith user being the sender, this probability is decided by attackers. HM indicates the maximum entropy of the system. Formation (1) indicates that the anonymity degree equals to the ratio of the system entropy with attackers' knowledge to the maximum system entropy.

Moser [46] adopted above quantifying method and ATC, and estimated the coin-mixing effect of mixers such as BitcoinFog.com, BitLaundry.com and Blockchain.info, and then found that he could find some clues about coin-mixing paths of the mixers except blockchain.info.

Quantifying the anonymity degree is a new method to estimate the coin-mixing effects, however, the present studies are too few and more attention and further research are required.

## 3. The Bitcoin Protocol and Sybil Attacks

*3.1. The Bitcoin protocol*

Bitcoin nodes communicate with each other via unencrypted TCP connections using port 8333 [47]. A Bitcoin wallet which does not accept incoming connections is known as the Client while others are called Bitcoin Nodes. Both the client and node save the copy of the IP addresses and ports of other clients and nodes. By default, they always keep 8 outgoing connections. If the number of outgoing connections falls below 8 they will reconnect until then number returns back to 8 entry nodes.

Both the client and node keep a record of other client's and node's penalty points. Penalty points are used as the basis of a disconnecting mechanism to avoid denial of service (DOS) attacks. When illegal blocks and transactions occur, the originating node will incur penalty points. Then, when the points total reaches 100, all connections from it will be rejected for 24 hours as a punishment.

When the client generates a new transaction, the command 'inv' is sent to the entry node. The entry node checks the transaction id in its own transaction database, if it exists, the id is disregarded, if not, it will send the command 'getdata' to request the contents of this transaction. The client replies with the command 'tx' as well as the transaction data or replies 'notfound' otherwise. Then the entry node verifies the transaction. If the transaction is not correct, it will return 'reject', if it is correct, the transaction will be transmitted to its entry nodes.

*3.2. Anonymous Network*

TOR [32] and I2P [33] are anonymous networks, which hide the real IP address for users and encrypt the transferred data, preventing original data from being exposing.

The key technologies are: ① route selection, the client decides the route randomly. As for TOR, Each node knows the identity of its previous node and next node, but it does not know the source IP and destination IP. The last node decrypts data and accesses the Internet in plain text. As for I2P, the original data was divided into several packages, and each package was transferred in a different route; ② data encryption, the original data was encrypted when across through the relay nodes, and then decrypted at the last node, and the relay nodes don't know the original data.

TOR and I2P are general anonymous networks, not specially serving for the Bitcoin network. In 2015, FBI closed 400 illegal websites on the TOR network, and it seems the security of TOR under serious threat.

*3.3. Sybil Attcack*

In active or passive ABPN, there are several attacks in which the new transaction can be linked to an IP address so that the attacker could find the relationship between the Bitcoin address and IP address.

**Bitcoin Protocol Sniffer**: In the Bitcoin protocol, the data is not encrypted so a well-formed sniffer could monitor all outgoing 'inv' commands to check whether the transaction id is likely to be a new transaction. Hence, we could obtain the relationship between the Bitcoin address and the IP address, and get the real identity with the help of the telecom operator's IP records.

*Sybil Attack*: Kaminsky [10] proposes the Sybil attack using a Bitcoin client to connect to all nodes in the Bitcoin network. The first source IP address of a new transaction is owned by the original sender. Koshy's [11] experiments, which had gathered 2,500,000 pairs of address and IP within 5 months, show that this method works, but there are three problems remaining: (1) Bitcoin via TOR hides the true IP address, (2) a large number of clients cannot be connected directly, (3) the same client owns different sessions, different IP addresses and different networks (anonymous and not anonymous), so it is difficult to link the transactions and IP addresses.

*Sybil attack plus entry nodes*: Biryukov [12] implemented a method that makes all Bitcoin nodes deny connections from TOR exit nodes in 2014. Meanwhile, they succeeded in detecting the entry nodes of a specified client. With these two tricks, they solved the first two problems of the Sybil attack. Those suspicious IP addresses collected in a Sybil attack include the IP address of the sender, IP addresses of entry nodes, and IP addresses of non-entry nodes. Through the delivery time of every mentioned IP addresses, we could probably then find the source IP address of a new transaction. Their results on a Bitcoin test network show that there is a 60% chance of identifying the source IP of a new transaction successfully using this method.

*Fake nodes attack:* Alex Biryukov [13] developed a TOR middle-man attack and 'Address cookies' to solve the third problem in 2015 via what could be called a fake Bitcoin nodes attack. This works by firstly, establishing a sufficient number of fake TOR exit nodes (the amount should reach 3% of all exit nodes in TOR network) and fake Bitcoin nodes (1,000 to 1,500). These fake nodes behave like normal nodes but they run code from the attackers. Then, address cookies aim to identify a certain client even if it uses different IP addresses, different sessions, and different networks.

### 3.4. Transaction remote release (TRR)

In the Bitcoin protocol, the only way that the attackers can connect the Bitcoin address with an IP address is in the process of spreading a new transaction. If we encrypt the new transaction and obfuscate the source IP of the sender then the attacker may not succeed.

Shentu etc. [34] proposed a new anonymization technology for Bitcoin, which denoted as Transaction Remote Release (TRR). A client encrypts a new transaction layer by layer, using the public key from different TRR nodes (Bitcoin nodes supporting TRR protocol). Then it establishes an independent connection to TRR nodes, one by one. When a TRR node receives data, it decrypts the data using its private key and then transmits the remaining data to the next node. The last TRR node releases the transaction to the Bitcoin network. Each node knows the identity of its previous node and next node, but it does not have access to the transaction content. Only the client and the last node know the content of the transaction, but the last node does not know the IP address of the client.

TRR doesn't require TOR or I2P, using public key to encrypt data, avoiding entry nodes of the Bitcoin nodes, and rendering the ABPN ineffective, thus provides strong anonymity for Bitcoin. However, TRR would require to modify the Bitcoin protocol, thus cannot be applied at once. Setting up a new blockchain and establishing a Transaction Delivery Network (TDN) are two possible avenues for further research. DarkNetspace [42] is an independent blockchain of crypto-currency, based on TRR technology to enhance the anonymity of currency transactions. TDN is an independent network based on TRR technology to distribute new transactions from any blockchain anonymously, supporting multi-currencies and multi-blockchains.

4. **Research directions and Prospects**

We think that there are several directions worth further research.

(1) There are some representative studies on ATC but not yet reaching the practical stage. A large number of stolen Bitcoins failed to be identified its owner, so it is necessary to set up a global identify database including Bitcoin addresses and users identifies. In addition, some stolen Bitcoins is difficult to be traced through ATC because they were mixed by mixers.

(2) Research on the anonymity of Bitcoin network is very limited, when the TOR denying and TOR middle-man were conducted successfully, the anonymity of the Bitcoin network was seriously under threat. TRR is a useful exploration, and more research work is expected.

(3) The security and practicability of the decentralized coin-mixing protocols have not estimated adequately. More deanonymization research is expected to attack decentralized coin-mixing protocols.

(4) Expecting that more cryptography algorithms would be applied on the Bitcoin system and would strength the security of privacy of Bitcoin, which are group signature, group blind signature, privacy sharing, homomorphic encryption, lattice cryptography and other algorithms.

(5) The research on the anonymity of Bitcoin is always qualitative. Quantifying research method and mathematical model are badly needed.

(6) Getting rid of the risk of currency loss for users is also an urgent subject for centralized coin-mixing.